\documentclass[a4paper,11pt]{article}

\usepackage{amsmath}
\usepackage{amssymb}
\usepackage{amsthm}
\usepackage{indentfirst}
\usepackage{graphicx}
\usepackage[utf8]{inputenc}
\usepackage[dvipsnames]{xcolor}
\usepackage{tabularx}
\usepackage{caption}
\usepackage{subcaption}
\usepackage{gensymb}
\usepackage[normalem]{ulem}

\usepackage{jheppub}
\hypersetup{colorlinks=true, linkcolor=blue, urlcolor=blue, citecolor=blue, linktocpage=true}

\begin{document}

\title{Rejecting noise in Baikal-GVD data with neural networks}

\author[a,b] {I. Kharuk,}
\author[a,c] {G. Rubtsov,}
\author[a,d] {G. Safronov}

\affiliation[a]{Institute for Nuclear Research of the Russian Academy of Sciences,
\\ 60th October Anniversary Prospect, 7a, Moscow, 117312, Russia}
\affiliation[b]{Moscow Institute of Physics and Technology,\\
Institutsky lane 9, Dolgoprudny, Moscow region, 141700, Russia}
\affiliation[c]{Laboratory of Cosmology and Elementary Particle Physics, Novosibirsk State University,\\ Novosibirsk, 630090 Russia}
\affiliation[d]{Joint Institute for Nuclear Research, 
\\ Joliot-Curie 6, Dubna, Moscow Region, 141980, Russia}

\emailAdd{ivan.kharuk@phystech.edu}

\abstract{Baikal-GVD is a large ($\sim$1 km$^3$) underwater neutrino telescope installed in the fresh waters of Lake Baikal. The deep lake water environment is pervaded by background light, which is detectable by Baikal-GVD's photosensors. We introduce a neural network for an efficient separation of these noise hits from the signal ones, stemming from the propagation of relativistic particles through the detector. The model has a U-net-like architecture and employs temporal (causal) structure of events. The neural network's metrics reach up to 99\% signal purity (precision) and 96\% survival efficiency (recall) on Monte-Carlo simulated dataset. We compare the developed method with the algorithmic approach to rejecting the noise and discuss other possible architectures of neural networks, including graph-based ones.}

\maketitle

\section{Introduction}
\label{sec:introduction}

Baikal-GVD is a large-volume water-based neutrino telescope aimed at studying the flux of high-energy cosmic neutrinos and searching for their sources \cite{belolaptikov2021neutrino,Baikal-GVD:2022fis}. The experiment is located in Lake Baikal, Russia, and, as of 2022, has an effective working volume of approximately 0.5 km$^3$ with respect to the high-energy neutrino-induced cascades. The telescope is targeted to reach the volume of 1 km$^3$ by 2030 and is currently the largest neutrino telescope in the Northern Hemisphere. The location of Baikal-GVD and IceCube  \cite{IceCube:2013low} experiments makes them complimentary, as their data combined allows for comprehensive full-sky astrophysical surveys. The four major neutrino telescopes, i.e. IceCube, Baikal-GVD, KM3NeT \cite{aiello2019sensitivity}, and ANTARES \cite{adrian2014searches}, cooperate within the Global Neutrino Network.

\begin{figure}
\center{\includegraphics[width=0.85\linewidth]{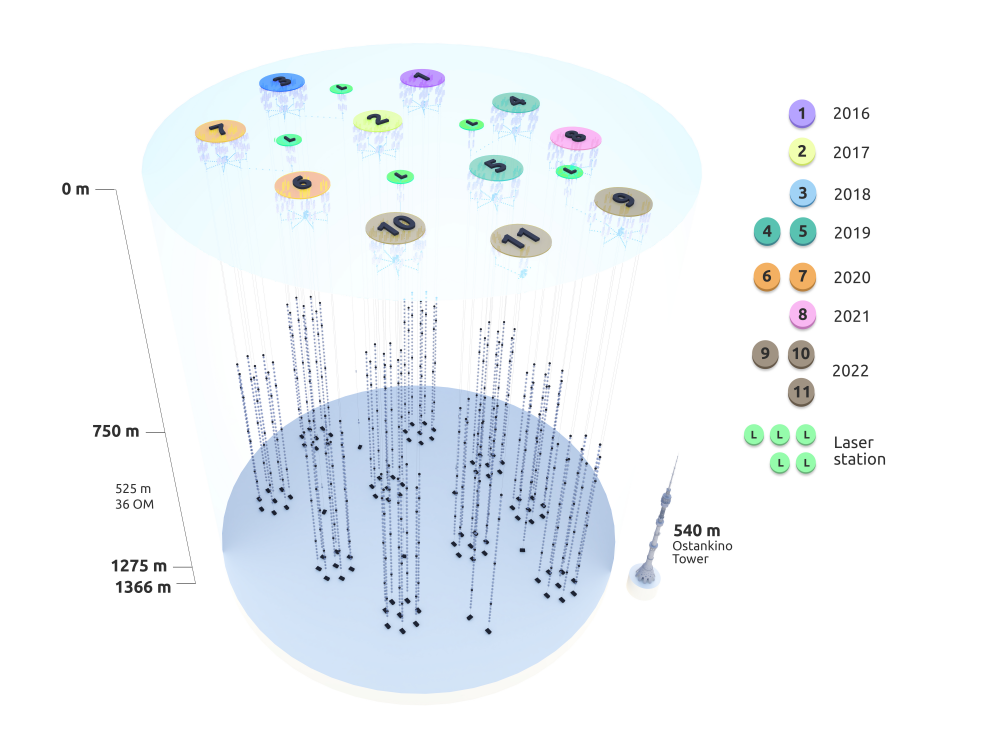}}\\
\caption{Schematic representation of the Baikal-GVD detector.}
\label{scheme_view}
\end{figure}

The basic components of Baikal-GVD's detector are \textit{optical modules} (OMs) accommodating 10-inch (25 cm) high quantum efficiency photomultipliers (Hamamatsu R7081-100). They are designed to register Cherenkov light produced by relativistic particles (originating from cosmic neutrinos and air showers) in the effective volume of the detector. OMs are carried by \textit{strings} (36 OMs per string) and are located at a depth of 750 to 1275 meters with a 15-meter spacing. The strings are organized into \textit{clusters} -- approximately regular heptagons with a string in the middle, see figure \ref{scheme_view}. In total, as of 2022, there are 11 clusters with an average distance of 300 meters between the clusters and an average cluster radius of 60 meters. 

\begin{figure}
\center{\includegraphics[width=0.45\linewidth]{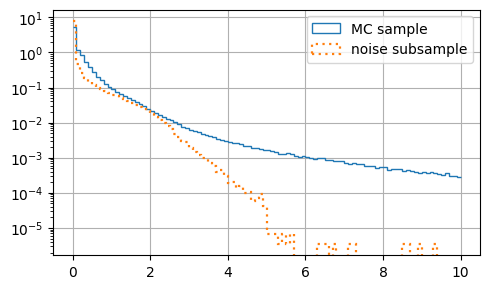}}\\
\caption{Normalized distributions of integral charges registered by OMs for all Monte-Carlo data (solid, blue) and noise subsample (dotted, orange).}
\label{qs_distr}
\end{figure}

As OMs are located underwater, they are subject to the natural luminescence of Baikal's water and to the photoemission of molecules and atoms in excited states \cite{avrorin2019optical}. Corresponding random, uncorrelated activations of OMs constitute background (\textit{noise}) hits in the data. The charge deposition spectrum of these activations is well studied and is presented in figure \ref{qs_distr}. Noise hits have the rate of 20--100 kHz (depending on depth and season), and their charge deposition is of the order of 1 photo-electron (p.e.). On average, noise hits constitute approximately 85--90\% of the data collected for the analysis. In what follows, we will refer to the non-noise hits as \textit{signal} ones.

One can suppress the noise hits by introducing a signal-level threshold around several p.e. This will, however, suppress some of the signal hits as well. This is undesirable, since the reconstruction algorithms would benefit from keeping as much signal hits as possible. Therefore, effective filtering algorithms are essential.

In this paper we present a neural network for rejecting background hits in a standalone cluster readout regime. The main features of the developed method, discussed in detail in the main part of the paper, are the following:
\begin{itemize}
    \item Representation of Baikal-GVD's data in a form that makes direct use of the causal structure of the events; 
    \item Introduction of multiplicative Gaussian noise for proper simulation of the fluctuations of OMs readings;
    \item The use of a custom loss function that reduces time residuals of the hits identified as signal ones by the neural network;
    \item Neural network is insensitive to the auxiliary hits, which are introduced to make data representation uniform; 
    \item Compared to the noise suppression algorithms developed by the Baikal-GVD collaboration \cite{belolaptikov2021neutrino,allakhverdyan2021efficient}, our neural network has better metrics and allows for a much faster data analysis.
\end{itemize}
The developed method can also be readily extended to process the multi-cluster data.

The paper is structured as follows. In section \ref{sec:MC} we describe the Monte-Carlo simulations of Baikal-GVD's data. In section \ref{sec:NN} and appendix \ref{sec:GNN} data representation and neural network's architecture are discussed. The results, including comparison with non-machine learning methods, are presented in section \ref{sec:results}. Finally, section \ref{sec:conclusion} concludes the paper.

\section{Monte-Carlo simulations}
\label{sec:MC}

Simulation of the data is performed via the Monte-Carlo method. Two types of incoming particles are considered: 1) muon neutrinos arriving from under the horizon, and 2) bundles of muons originating from the cosmic air showers. The energy spectrum and incoming directions of arriving particles are chosen to coincide with the ones expected in the experiment, see \cite{allakhverdyan2021measuring,stasielak2021high} for details. The procedure takes into account photon scattering in Baikal's water and full simulation of cosmic air showers' evolution using QGSJET II-03 \cite{kalmykov1993nucleus} and CORSIKA \cite{heck1998corsika}. The noise rate and charge distribution are simulated so as to mimic the experimentally expected detector conditions. 

The triggering condition for identifying an \textit{event} is the following: two adjacent OMs, within the 100 ns time window, register signals that are at least 4.5 and 1.5 p.e. If this condition is fulfilled, the data from all OMs exceeding the signal-level threshold (0.3 p.e.) is collected for further analysis.

Registered signals (waveforms) are approximated by discrete hits (pulses). Each of the hits is characterized by the following physical observables:
\begin{enumerate}
\item Time at which the hit was registered;
\item Integral charge (in p.e.) registered by OM;
\item Maximal amplitude of the registered signal.
\end{enumerate}

Additionally, in Monte-Carlo simulated data each hit is supplemented by a flag indicating its origin. We reduced these flags to a binary mask indicating whether a given hit is a signal or noise one. This allows us to pose the problem of rejecting the background noise as a supervised segmentation task, i.e. predicting whether a given hit in an event belongs to a noise or a signal class.

Monte-Carlo simulations do not properly model OMs' response when the integral registered charge is larger than 100 p.e. (due to OM saturation effects). To account for this fact, we set integral charge to be 100 p.e. for all OMs exceeding this value. For consistency, it is assumed that the same procedure would be applied for the experimental data as well.
 
For training the neural network, we used neutrino-induced and cosmic ray-induced events. To avoid bias related to different signatures of these two types of events, we took an equal number of samples of both types, approximately $ 10^7 $ entries each. It was observed that neural network's metrics on both types of events are similar, and that changing the proportion does not significantly affect the results. The data has been split into train, validation, and test sets in proportion 8/1/1. 

\section{Neural network architecture}
\label{sec:NN}

\subsection{Data representation}
\label{sec:data_rep}

Performance of a neural network strongly depends on a way the data is represented (sequence, image, graph). Hence it is of primary importance to choose the most efficient representation. For Baikal-GVD, we have three possible options:

\begin{enumerate}
\item Organize data in a three-dimensional array mimicking geometrical structure of a cluster;
\item Make a sequence by ordering hits according to their activation times, thus exploiting temporal (causal) structure of the events;
\item Consider each event as a graph.
\end{enumerate}
In what follows we refer to these options as geometrical, temporal, and graph representations correspondingly. Variations and combinations of these representations have been applied for data analysis in physical experiments \cite{drielsma2021clustering,bister2021identification,kalashev2022deep,domine2021point,domine2020scalable}, including neutrino telescopes \cite{aiello2020event,choma2018graph,huennefeld2017deep,huennefeld2019reconstruction}. Below we comment on each of the possibilities one by one, highlighting their pros and cons for our task. In each case, the information on hits is given by the following 5 parameters:
\begin{enumerate}
\item[1)] Integral charge registered by OM during a pulse;
\item[2)] Time the hit was registered at;
\item[3-5)] $x$, $y$, and $z$ coordinates of the corresponding OM ($xy$ plane is parallel to the water level, $z$ axis is perpendicular to it).
\end{enumerate}

For the geometrical representation, we have introduced a 3$\times$3$\times$36 array mimicking geometrical structure of a cluster (one of the 1$\times$1$\times$36 ``strings'' is auxiliary). This representation has a drawback -- one can store only one OM activation per event, which might result in a loss of the signal hits. This can be resolved by introducing an additional time axis, similarly to \cite{aiello2019sensitivity}. However, this is highly memory-demanding, and we decided to keep only the activation with the highest integral signal per OM. It was verified that in this way we loose only a negligible fraction of the signal hits.

To analyze the data, we used a ResNet-like neural network \cite{he2016deep}. It is a convolution-based neural network with residual connections, which is often used for image classification and segmentation. However, its performance turned out to be 1-2\% worse than that of the best solution. Combined with the abovementioned drawback, we decided not to develop this approach further.

For the graph representation of the data, we have developed a graph neural network of the message passing type \cite{gilmer2017neural} with a modification of the attention mechanism \cite{velivckovic2017graph}. We have also implemented a non-standard message parsing algorithm utilizing temporal ordering of the activations. The details on data representation and graph updating protocol are given in the appendix \ref{sec:GNN}. The performance of this neural network is similar to that of the best solution, but it was slower and more memory-demanding.

The best results were obtained by utilizing the temporal representation of the data. Namely, for a given event, all of the hits are ordered according to their activation times.\footnote{Note that the temporal representation of the data can be considered as a special limit of the geometrical representation with time slicing -- when each time slice contains precisely one hit.} In such order, signal hits form clusters, which allows the neural network to efficiently identify them. We believe that this is the underlying reason why temporal representation turned out to be the best one for our task.

The number of OMs activated in an event (the \textit{``length''} of an event) follows some probability distribution, and thus is not fixed. For this reason we allowed for the input to the neural network to have arbitrary \textit{length} dimension. In each data batch, however, the shape of the events (as an in-memory arrays) must be the same. To fulfill this requirement, we padded all of the events to the same length -- the maximal one in a given dataset ($ N_{max} $).\footnote{Such convention is useful as Monte-Carlo and experimental data may have different $N_{max}$. Moreover, $N_{max}$ can change after extending Monte-Carlo simulations or getting new experimental data.} The corresponding auxiliary hits were filled with zeros and stacked to the data on the right (after the last real hit in an event). We also made a mask channel to keep track of the auxiliary hits. This provides the neural network with the information that the corresponding hits are special and should be processed differently. Thus, each event was represented by a $N_{max} \times (5+1)$ array.

One can prescribe physically meaningful data to the auxiliary hits. For example, as if the corresponding auxiliary OM was located outside the detector and registered zero integral charge at some late time (after the last hit in an event). Our tests and results, described in section \ref{sec:results}, have shown that to be sub-optimal. Namely, making the neural network explicitly insensitive to the auxiliary hits improves the metrics and makes it robust to varying length of events. The way how one can ensure the insensitivity to the auxiliary hits is described in detail in the next section.

\subsection{Neural network architecture}
\label{sec:nn_arch}

\begin{figure}
\center{\includegraphics[width=0.99\linewidth]{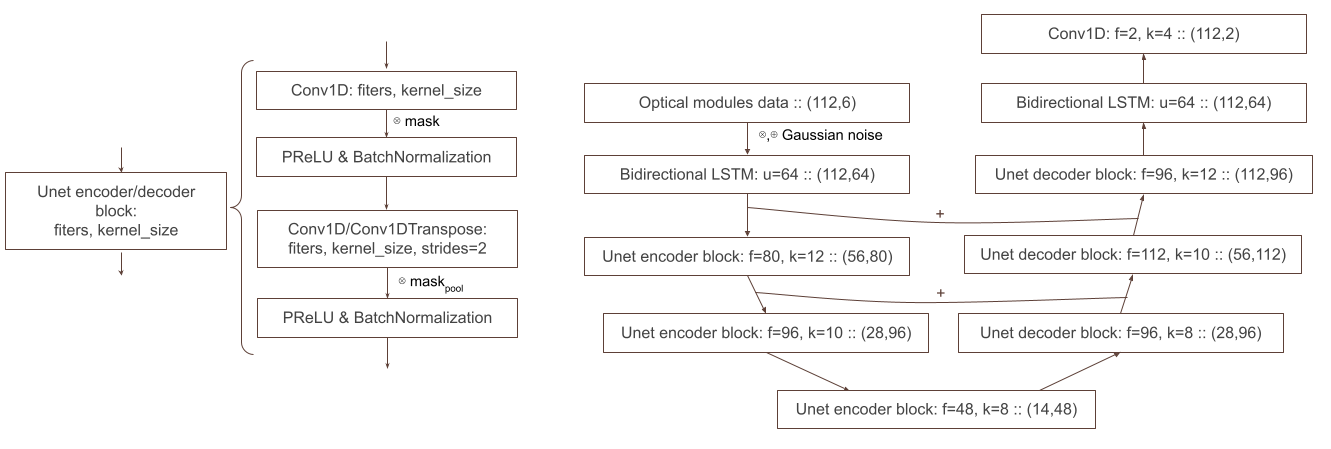}}\\
\caption{Architecture of the neural network. Arrows show the data flow, plus sign indicates concatenation of the data. $N_{max}$ is set to 112 for illustration purposes.}
\label{nn_arch}
\end{figure}

One of the neural network architectures known to be well suited for segmentation tasks is the U-net \cite{domine2021point,long2015fully,ronneberger2015u,zhou2018unet++}. Its power comes from the fact that it utilizes both global and local information. Namely, convolutions without dimensionality reduction (U-net encoder blocks, figure \ref{nn_arch}) provide local information for a given OM from the neighboring cells. Convolutions with dimensionality reduction followed by transposed convolutions (U-net decoder blocks) have significantly larger receptive field. This provides the neural network with the information on global features of an event as well. Combining these two types of information allows for an effective segmentation of the data. This was our motivation for choosing the U-net as a basic neural network architecture.

We found it beneficial to wrap (i.e. add a layer prior and after) the U-net block with bidirectional long short-term memory (LSTM) cells working in a sequence-to-sequence regime. Recurrent units, such as LSTM, are designed for analyzing sequences \cite{aab2021extraction,bahdanau2014neural,sutskever2014sequence}. Thus, by adding these additional layers, we made use of the fact that the data is provided in a form of a sequence.


The last layer of the neural network is a convolution block with softmax activation function. For each hit, its output is a pair of numbers, $(c,1-c)$, $ c \in [0;1] $, where $c$ represents neural network's confidence that a given hit belongs to the signal class. By setting a threshold $ c=c_0 $, all hits can be binarely split into signal and noise ones. Further we refer to the hits identified as signal ones by the neural network after such a split as \textit{NN-selected} hits. 

To make neural network insensitive to the auxiliary hits, we have made the following:
\begin{itemize}
    \item All calls to LSTM cells are supplemented with an additional \textit{mask} parameter, with our mask channel as the corresponding argument.\footnote{This is a part of the standard TensorFlow API.} This ensures that only non-auxiliary hits are processed by these cells.
    \item The outputs of all LSTM cells and convolutions are multiplied by the proper mask, which depends on the data dimensionality (length). For the initial length, this mask coincides with our initial mask channel. For the data which dimensionality has been reduced $2^n$ times, via series of convolutions with strides 2, the proper mask is obtained by applying \textit{GlobalPooling} layer, with strides 2, $n$ times to the initial mask. This procedure ensures that the auxiliary part of the data at any step is filled with zeros and does not affect the non-auxiliary part.
    \item The predictions after the last convolution layer are modified so that auxiliary hits are always assigned to the noise class. As a result, the neural network does not need to learn to make the correct prediction for the auxiliary hits, and they do not affect the backpropagation.
\end{itemize} 

To avoid overfitting (in particular, to Monte-Carlo simulations, as opposed to real experimental data), we have introduced a Gaussian noise into the training data. It comes in two forms:


\begin{itemize}
    \item \textit{Additive noise}. We add to the data a random Gaussian noise with the following standard deviations: 0.3 p.e. for the registered integral charge, 6 ns for the OMs' activation time, 20 cm for $x$ and $y$ coordinates of OMs, and 2 cm for the $z$ coordinate of OMs. These values have been chosen so that they are approximately 1,5 time larger than the expected systematic experimental uncertainties, which allows us to account for the modelling uncertainties as well.
    \item \textit{Multiplicative noise}. To properly model the fluctuations of the integral charge registered by OMs when it is bigger than 10 p.e., we have introduced a multiplicative noise. Namely, we add to the corresponding data channel the Gaussian noise centered at zero with standard deviation of 10\% of the signal level.
\end{itemize}

For training the neural network, we have introduced a special loss function that penalizes it for identifying hits with big time residuals as signal ones. Time residual is the difference between the expected time of arrival of the Cherenkov light, and the actual one.\footnote{This difference in time originates due to the fluctuations of light propagation in Baikal's water.} Rejecting hits with big time residuals is important as even a single such hit will worsen the accuracy of the reconstruction algorithm for neutrino's incoming angle. 

For an individual hit, the loss function is given by the formula
\begin{equation} \label{loss_func}
    Loss(y_{true},y_{pred},t_{res},t_{th}) = Loss_{entropy}(y_{true},y_{pred}) + c \cdot Loss_{trespen}(y_{pred},t_{res},t_{th}) \;,
\end{equation}
where $y_{true}$ is the true label, $y_{pred}$ is neural network's prediction, $Loss_{entropy}$ is a binary cross-entropy loss function, $c$ is a numerical coefficient, $t_{res}$ is hit's time residual, $t_{th}$  is the threshold value, and $Loss_{trespen}$ is given by the formula
\begin{equation}
    Loss_{trespen} = \begin{cases}
    0, ~ &\text{if } |t_{res}|< t_{th} \;, \\
    y_{pred}^{sig} \cdot |t_{res}| &\text{if } |t_{res}| \geq t_{th} \;,
    \end{cases}
\end{equation}
where $y_{pred}^{sig}$ is neural network's confidence that the given hit is a signal one. For an event, the loss is averaged over all hits. Empirically we found $c=0.1$ and $t_{th}=20$ ns as optimal values. This loss function allowed us to reduce the mean and dispersion of time residuals of the NN-selected hits by 50\% and 10\% accordingly.

Full architecture of the neural network is presented in figure \ref{nn_arch}. The code for reproducing neural network's architecture and the used loss function can be found at \url{https://github.com/ml-inr/Baikal-GVD-noise-rejection}.

We used TensorFlow \cite{abadi2016tensorflow} to implement the neural network. For training, we employed Adam optimizer \cite{kingma2014adam} with a dynamical learning rate and used early stopping to avoid overfitting. The activation functions were chosen as follows: PReLU \cite{he2015delving} in the U-net block, hyperbolic tangent in LSTM cells, and softmax in the last layer of the neural network. The training takes approximately 36 hours on NVIDIA RTX 3090.\footnote{The training converges after 12-18 hours, at later times the metrics become only slightly better.}

We have also tried training neural networks based solely on LSTM cells or convolution blocks without dimensionality reduction. Their performance was slightly worse.

\section{Results}
\label{sec:results}

Neural network's performance has been evaluated on the test dataset -- an equal-part mixture of neutrino-induced and muon bundle-induced events. 

\begin{figure}
\center{\includegraphics[width=0.3\linewidth]{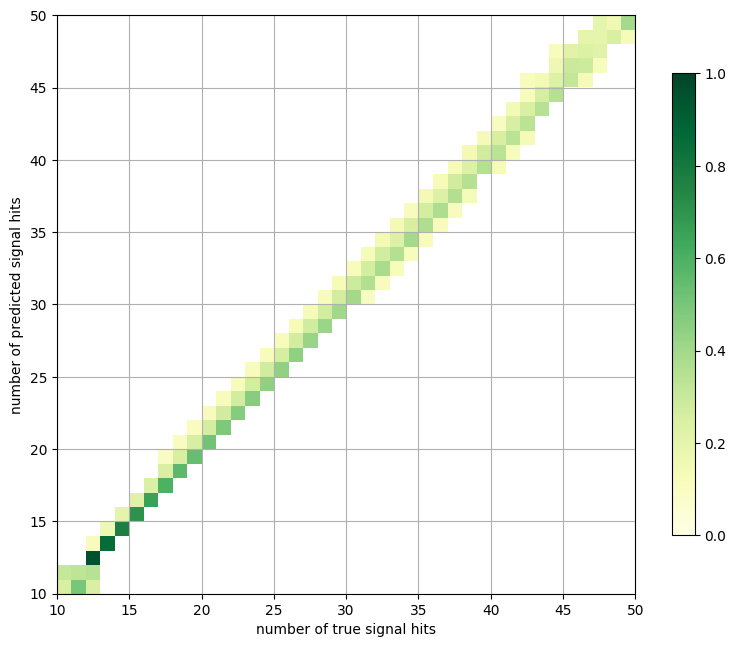}}\\
\caption{Response matrix for the number of reconstructed signal hits for the classification threshold of 0.7. The number of true and reconstructed signal hits are given on horizontal and vertical axes correspondingly.}
\label{resp_matrix}
\end{figure}

We start by considering neural network's response matrix for the number of reconstructed signal hits, figure \ref{resp_matrix}. The corresponding plot, as it should be for a well-trained neural network, can be approximated by a straight line. As the number of signal hits and event energy are correlated, this also implies that neural network's predictions are reliable in the whole range of the energy spectrum.

To quantitatively estimate neural network's performance, we used the following metrics: signal purity (precision), survival efficiency (recall), and the dispersion of time residuals of the NN-selected hits.\footnote{Time residuals' mean is around 0.1-0.3 ns (depending on the identification threshold), and is not as important as theirs dispersion.}

\begin{figure}
     \centering 
     \hspace*{\fill}
     \begin{subfigure}[b]{0.32\textwidth}
         \centering
         \includegraphics[width=\textwidth]{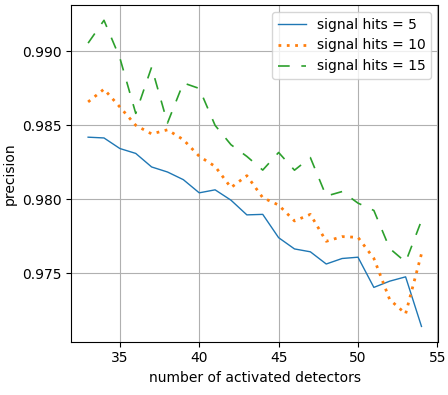}
         \caption{}
     \end{subfigure}
     \hfill
     \begin{subfigure}[b]{0.32\textwidth}
         \centering
         \includegraphics[width=\textwidth]{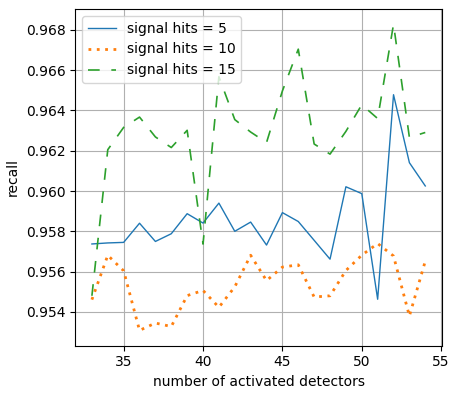}
         \caption{}
     \end{subfigure}
     \hspace*{\fill}
        \caption{Plots of the dependence of precision (a) and recall (b) as functions of the number of hits registered in an event. Blue solid, orange dotted, and green dashed curves correspond to events with 5, 10, and 15 signal hits correspondingly.}
        \label{metrics_num_hits}
\end{figure}

Figure \ref{metrics_num_hits} illustrates the dependence of precision and recall on the number of registered hits in an event for the identification threshold of 0.5. As it can be expected, the neural network performs slightly better on events with a higher fraction of signal hits.

\begin{figure}
     \centering
     \hspace*{\fill}
     \begin{subfigure}[b]{0.32\textwidth}
         \centering
         \includegraphics[width=\textwidth]{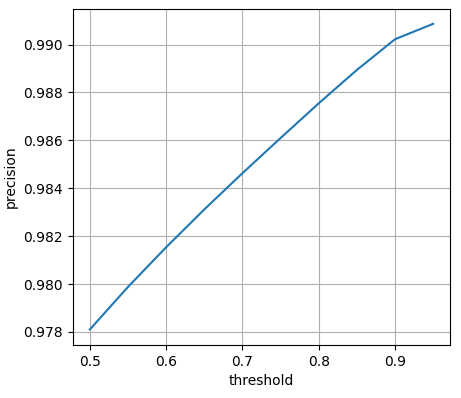}
         \caption{}
     \end{subfigure}
     \hfill
     \begin{subfigure}[b]{0.32\textwidth}
         \centering
         \includegraphics[width=\textwidth]{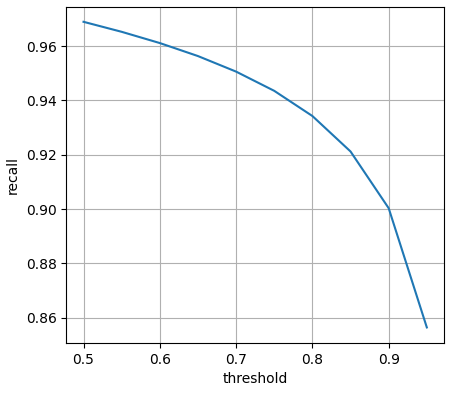}
         \caption{}
     \end{subfigure}
     \hfill
     \begin{subfigure}[b]{0.32\textwidth}
         \centering
         \includegraphics[width=\textwidth]{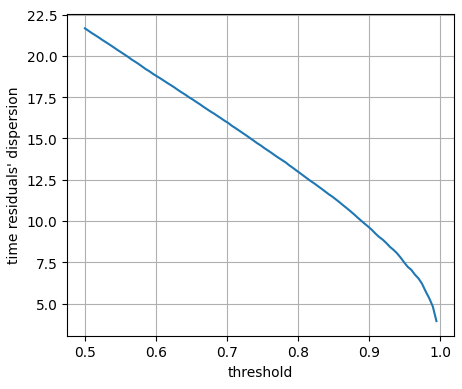}
         \caption{}
     \end{subfigure}
     \hspace*{\fill}
        \caption{Dependence of precision (a), recall (b), and time residuals' dispersion (c) as the function of the identification threshold value.}
        \label{metrics_threshold}
\end{figure}

The dependence of the metrics on the identification threshold is shown in figure \ref{metrics_threshold}. We estimated the errors of metrics' evaluation by training several neural networks and comparing their predictions. The resulting errors for precision, recall, and dispersion of time residual were found to be 0.1\%, 0.2\%, and 4\% accordingly, and are independent of the threshold value. 

\begin{figure}
     \centering
     \hspace*{\fill}
     \begin{subfigure}[b]{0.32\textwidth}
         \centering
         \includegraphics[width=\textwidth]{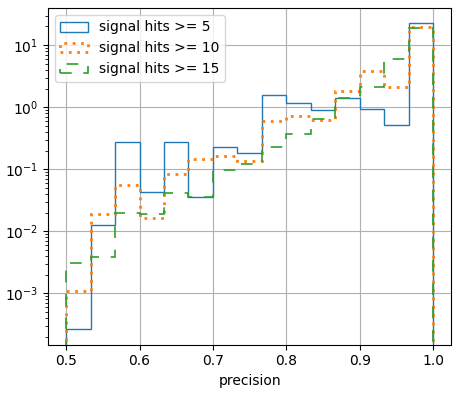}
         \caption{}
     \end{subfigure}
     \hfill
     \begin{subfigure}[b]{0.32\textwidth}
         \centering
         \includegraphics[width=\textwidth]{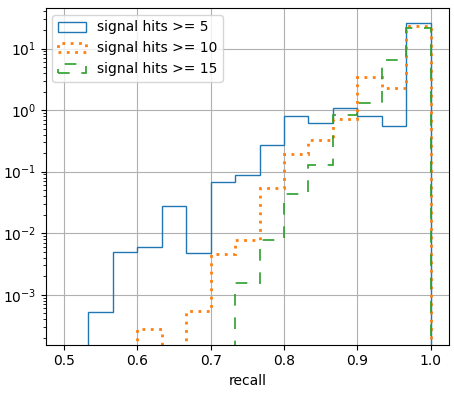}
         \caption{}
     \end{subfigure}
     \hfill
     \begin{subfigure}[b]{0.32\textwidth}
         \centering
         \includegraphics[width=\textwidth]{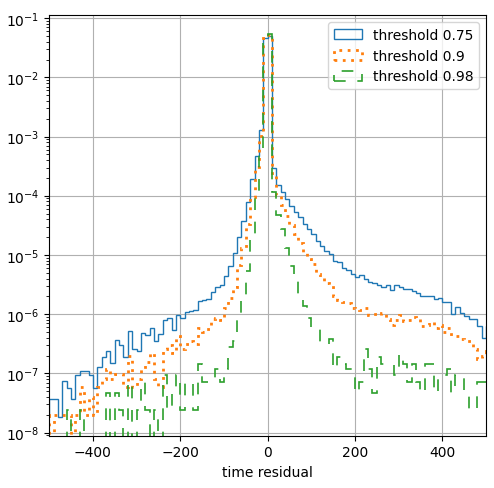}
         \caption{}
         \label{tres_dist}
     \end{subfigure}
     \hspace*{\fill}
        \caption{Distribution of the metrics for individual events. 
        (a) and (b) display distributions of precision and recall for the identification threshold of $0.5$; blue solid, orange dotted, and green dashed curves correspond to the events with at least 5, 10, and 15 signal hits correspondingly. 
        (c) displays the distribution of time residuals for various values of the identification threshold.}
        \label{metrics_dists}
\end{figure}

The distribution of the metrics for individual events is shown in figure \ref{metrics_dists}. As it can be seen from the graphics, recall's dispersion strongly depends on the number of signal hits in an event. Figure \ref{tres_dist} displays the distribution of time residuals for the NN-selected hits for various values of the identification threshold. 

We observed that making the neural network insensitive to the auxiliary hits reduces the dispersion of time residuals by approximately 5\%. Introduction of the loss function described in the previous section, Eq. \ref{loss_func}, reduces it further by $\sim$10\%. These modifications combined have only slightly affected precision and recall, making them $\sim$0.2\% worse. Introduction of the Gaussian noise worsened precision, recall, and dispersion of time residuals by approximately 0.3\%, 0.6\%, and 10\% correspondingly. 

Neural network's performance has been compared to the algorithmic approach to rejecting the noise \cite{belolaptikov2021neutrino,allakhverdyan2021efficient}. For this comparison we have chosen a subset of atmospheric neutrino events on which the method in question identifies at least 8 signal hits on at least 2 strings. These filters are standard for the algorithmic approach.

First, we compared the metrics. The algorithmic approach yielded 95\% precision, 95\% recall, and 5 ns dispersion of time residuals. For rejecting the noise with the neural network, we fixed the classification threshold value so that the dispersion of time residuals is the same as in the algorithmic approach. This yielded 99.5\% precision and 96\% recall, which is significantly better. 

\begin{figure}
     \centering
     \hspace*{\fill}
     \begin{subfigure}[b]{0.42\textwidth}
         \centering
         \includegraphics[width=\textwidth]{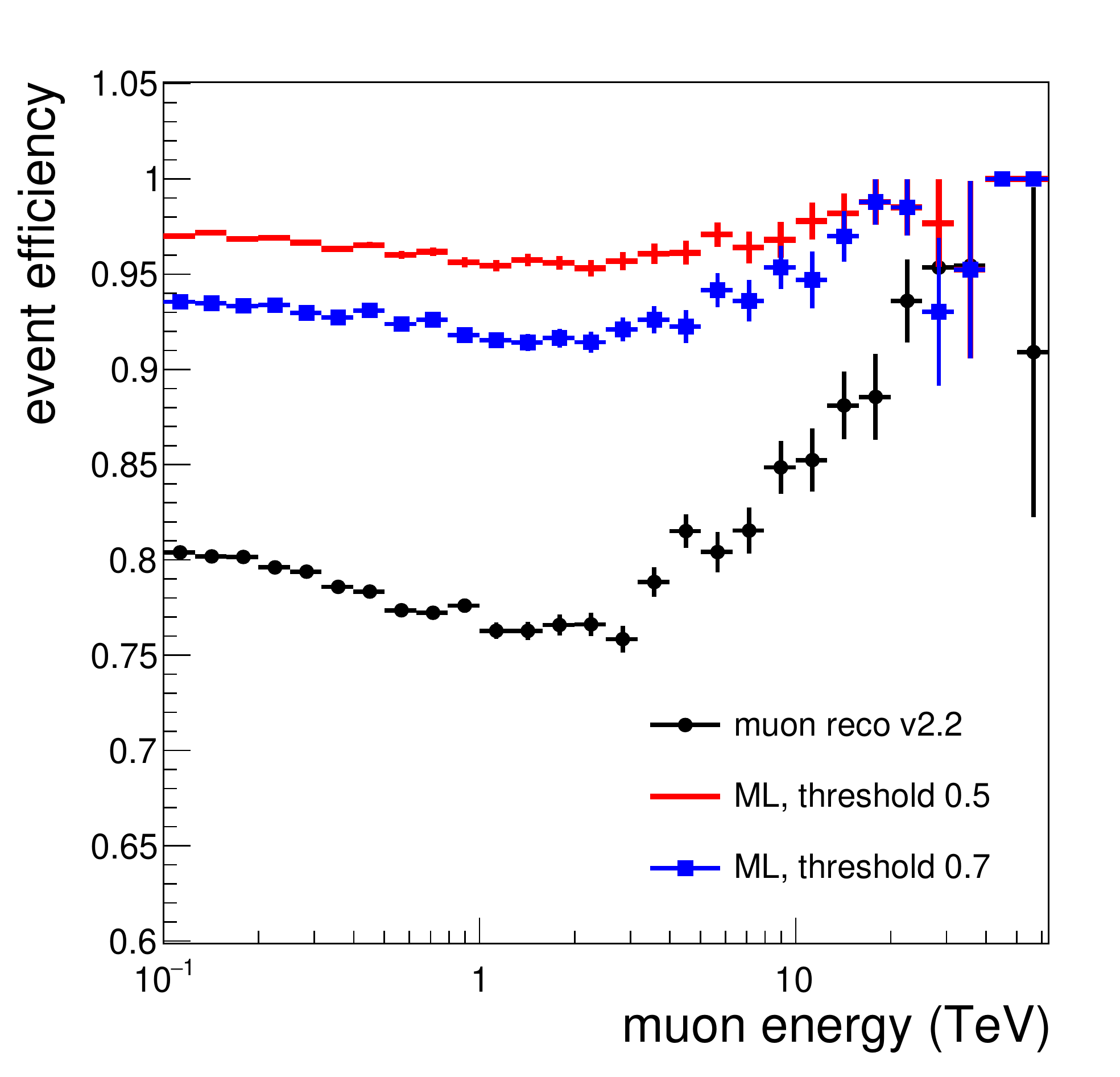}
         \caption{}
         \label{algo_vs_ml_efficiency}
     \end{subfigure}
     \hfill
     \begin{subfigure}[b]{0.42\textwidth}
         \centering
         \includegraphics[width=\textwidth]{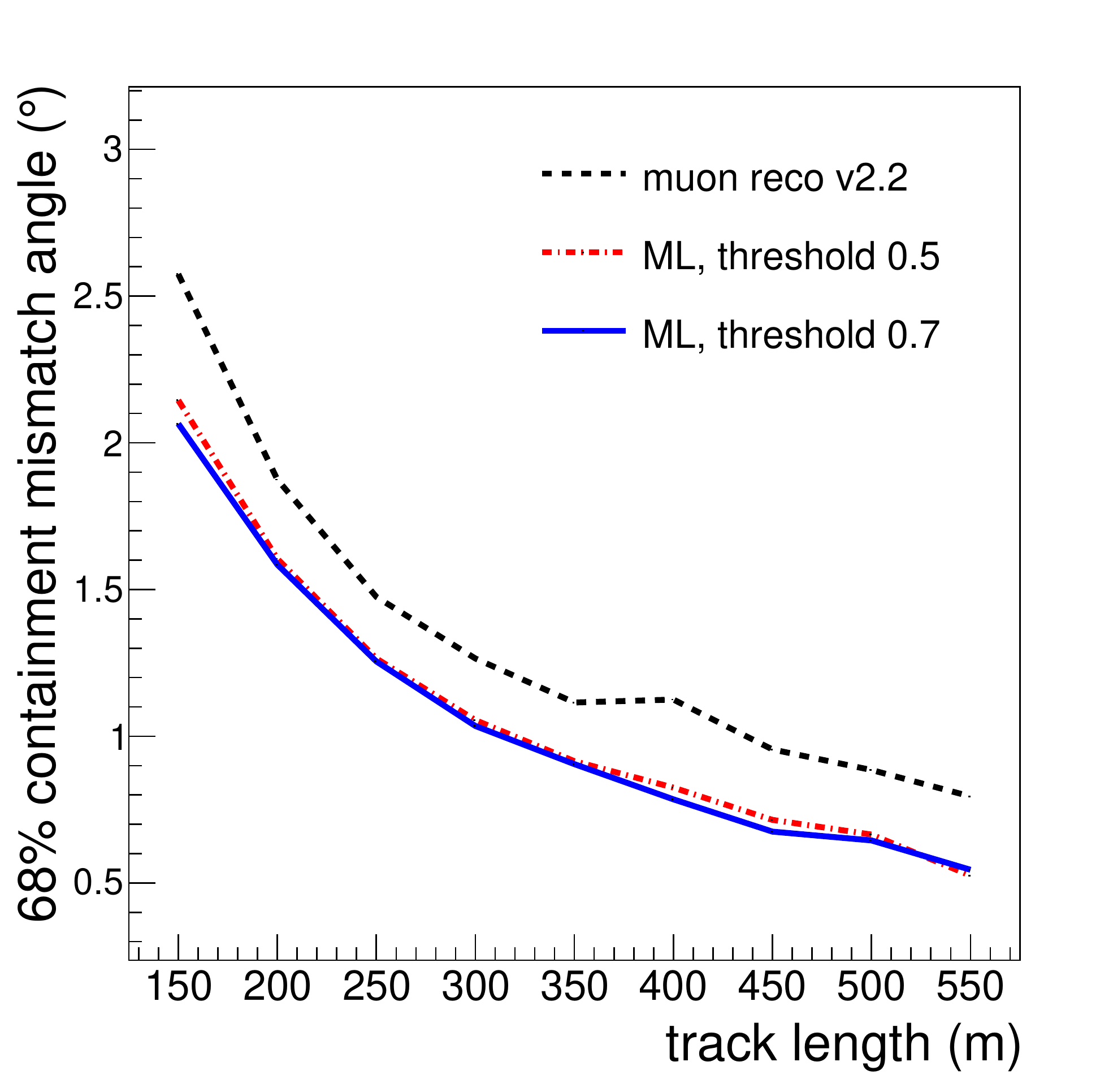}
         \caption{}
         \label{algo_vs_ml_resolution}
     \end{subfigure}
     \hspace*{\fill}
        \caption{ 
        (a) Comparison of event efficiency for algorithmic and machine learning approaches. Black dots show the performance of the algorithmic approach, blue squares and red dashes depict neural network performance for the identification thresholds of 0.5 and 0.7 correspondingly. Vertical bars indicate the statistical error of event efficiency estimation.    
        (b) Comparison of 68\% containment mismatch angle resolution for events filtered from noise hits by the algorithmic approach (black dotted line) and by the neural network (red dash-dotted and solid blue lines, corresponding to identification thresholds of 0.5. and 0.7).}
\end{figure}

Second, we studied the efficiency of event identification. By definition, the event efficiency is the ratio of the number of events in which the method in question identifies at least 8 signal hits on at least 2 strings to the true number of such events. As one can see from figure \ref{algo_vs_ml_efficiency}, neural network performance is better. 

Finally, the data filtered from noise hits by the neural network was used to reconstruct neutrino's incoming direction. The resulting angle resolution is compared to that of the standard algorithmic procedure in figure \ref{algo_vs_ml_resolution}. Using the neural network for rejecting the noise allows to improve the angle resolution by approximately 0.5\degree .

At the end of this section we would like to note that the usage of neural network allows for a much faster data analysis (up to $10^3$ times faster when data is processed in batches).

\section{Discussion and conclusion}
\label{sec:conclusion}

We have presented the neural network for rejecting noise in Baikal-GVD's data. The main features of the developed method are the following:
\begin{itemize}
    \item The usage of temporal (causal) representation of Baikal-GVD's data, which was found to be the optimal one;
    \item Utilization of the loss function that reduces time residuals of the hits identified as signal ones by the neural network;
    \item Insensitivity of the neural network to the auxiliary hits, which improved its robustness;
    \item Modeling of the fluctuations of the integral charge registered by OMs via a multiplicative Gaussian noise.
\end{itemize}
The neural network shows better metrics than the standard algorithmic approach (4\% and 1\% improvements in precision and recall, accordingly), and allows for a much faster data analysis (up to a factor of $10^3$).

The techniques developed in the paper may be useful for data analysis in other experiments. For example, most of the machine-learning based techniques used in astrophysics utilize geometrical representation of the data \cite{aiello2020event,huennefeld2017deep,huennefeld2019reconstruction,ivanov2020using}. As our results suggest, temporal (causal) representation of the data may yield better results not only for signal-noise separation, but for event reconstruction as well.

The developed method for analyzing events using a temporal representation of the data can be generalized to the case when input data are raw impulses (waveforms) registered by OMs. Namely, let the information on each hit be given by: (\textit{i}) a raw impulse (of a fixed length) and the corresponding (\textit{ii}) starting time of OM's recordings and (\textit{iii}) its coordinates. Then, using (\textit{ii}), the hits can be time-ordered. Further one should introduce a (sub-)neural network, an \textit{encoder}, that takes as input raw impulses and outputs their encodings.\footnote{These encodings can be thought of impulse characteristics, which were found to be the optimal ones for a given task. Note that encoder is not pre-trained and does not have its own loss function.} By applying the encoder to each of the raw impulses, the latter can be reduced to (\textit{i'}) a set of their encodings. Subsequent processing of the data, i.e. of (\textit{i'}), (\textit{ii}), (\textit{iii}), follows the algorithm described in the main part of the paper. Such procedure, for example, has been applied for the data analysis of the Telescope Array experiment \cite{kalashev2022deep}.

This work constitutes the first step towards introducing a full cycle of machine learning-based data analysis for the Baikal-GVD experiment. Namely, we are developing neural networks for an effective separation of neutrino-induced and muon bundle-induced events, their energy and arrival direction reconstruction. It is expected that the introduction of these neural networks will further improve data analysis quality.

\acknowledgments

The work is supported by the Russian Science Foundation grant number 22-22-20063.

\appendix

\section{Graph neural network}
\label{sec:GNN}

Graph neural networks have been successfully applied for solving various tasks \cite{adrian2014searches,drielsma2021clustering,bister2021identification,gilmer2017neural}, including data analysis of neutrino telescopes \cite{choma2018graph,reck2021graph}. This motivated us to develop a graph neural network for the purpose of rejecting the noise. 

We considered OMs as vertices of a graph. Each vertex has 6 parameters described in the main part of the paper. We assumed the initial graph to be fully connected. This is motivated by the fact that it is \textit{a priori} unknown which OMs are relevant for determining whether a given hit is a signal one or not. A simpler approach is to connect only $ k $ nearest neighbors, chosen according to the temporal ordering of OMs. This yields comparable yet slightly worse results.

We have chosen our neural network to be of the message passing type as they have most general, and hence powerful, architecture. Let us explain the data analysis protocol step by step, following the data flow in the neural network. Unless otherwise stated, the activation function of a layer is the scaled exponential linear unit. 

1) \textit{Nodes encoding}. At the first step of data analysis, we update the encodings of the vertices. This is done via a dense layer with 48 units, which updates each of the vertices encodings independently. 

2) \textit{Messages creation.} Further, we prepare messages between each pair of the vertices. This is done in a usual manner -- by a dense layer with 48 units that takes as input the encodings of two nodes and outputs a message from one node to another.
  
3) \textit{Attention mechanism}. At the next step, the obtained messages are multiplied by the corresponding values of the adjacency matrix. The idea is that we consider adjacency matrix as the attention mechanism, which leaves only relevant messages. The updating rule of the adjacency matrix is given below. 

4) \textit{Message parser}. To obtain a single message for a vertex, we order all messages sent to a given vertex according to the activation times of the corresponding OMs (temporal ordering). This sequence is passed via a convolutional layer (48 filters, kernel size 5, strides 2), followed by a maxpooling layer (strides 2). The motivation behind this procedure is to make use of a physically meaningful ordering of the vertices -- the temporal one. To obtain a single message, we sum the processed messages and divide the result by the sum of values in the adjacency matrix at the corresponding raw.

5) \textit{Nodes updating}. Further, vertices encodings are updated using a dense layer with 48 units. This layer takes as input current encoding of the vertex and the relevant message.

6) \textit{Adjacency matrix updating}. Finally, we update the adjacency matrix. Namely, we form all possible pairs of the vertices and pass them through a dense layer with 48 units and sigmoid activation function. The resulting output is interpreted as the mutual relevance of the cells. Such interpretation is supported by the fact that, upon examination of the adjacency matrix, it tends to have non-zero entries only between signal hits. Note that at the 2nd step we consider all possible pairing of the vertices, independently of the values of the adjacency matrix.

The described steps of data analysis constitute a single layer of our graph neural network. To get best results, we stacked 4 such layers on top of each other. The output of the last layer is passed through a dense layer with one output unit and the sigmoid activation function. This yields the neural network's confidence on whether a given hit is a signal one. 

We believe that the reason why the performance of the graph neural network wasn't better than that of the convolutional one is the following. Graph neural networks allow one to work with data that can be represented in a form of a graph. For the case at hand, this is not needed, as there is a useful and physically meaningful representation of the data -- temporal one. This allows neural network to understand the structure of the data, yielding graph representation unnecessary.

\bibliography{ref_baikal_sig-noise}

\end{document}